\documentclass[12pt]{article}

\usepackage{times}
\usepackage{braket}
\usepackage{graphicx}
\usepackage{amsmath}
\usepackage{mathrsfs}
\usepackage{xcolor}
\usepackage{longtable}
\usepackage{amsfonts}
\usepackage{tikz}
\usepackage[colorlinks,linkcolor=blue,citecolor=blue,anchorcolor=blue]{hyperref}
\usepackage{bm}
\usepackage{ulem}
\usepackage{mdframed}
\usepackage{amssymb}
\newcommand{\Yb}{$\mathrm{^{171}Yb}$~}

\usepackage{CJKutf8}

\newenvironment{sciabstract}{%
\begin{quote} 
 \bf}
{\end{quote}}

\title{Improving Metrology with Quantum Scrambling}

\author{Zeyang Li\begin{CJK*}{UTF8}{gbsn} (李泽阳)\end{CJK*},$^{1}$ Simone Colombo,$^{1}$ Chi Shu,$^{1,2}$ Gustavo Velez,$^{1,3}$\\
Sa\'{u}l Pilatowsky-Cameo,$^{4}$ Roman Schmied,$^{5}$ Soonwon Choi,$^{4}$\\ Mikhail Lukin,$^{2}$ Edwin Pedrozo-Pe\~{n}afiel,$^{1}$
Vladan Vuleti\'{c}$^{1\ast}$  \\
\\
\normalsize{$^{1}$Department of Physics, MIT-Harvard Center for Ultracold Atoms,}\\
\normalsize{Research Laboratory of Electronics, Massachusetts Institute of Technology,}\\
\normalsize{ Cambridge, Massachusetts 02139, USA}\\
\normalsize{$^{2}$Department of Physics, Harvard University,}\\
\normalsize{Cambridge, Massachusetts 02138, USA}\\
\normalsize{$^{3}$Department of Electrical Engineering and Computer Science,}\\
\normalsize{ Massachusetts Institute of Technology,}\\
\normalsize{Cambridge, Massachusetts 02139, USA}\\
\normalsize{$^{4}$Department of Physics, Massachusetts Institute of Technology,}\\
\normalsize{ Cambridge, Massachusetts 02139, USA}\\
\normalsize{$^{5}$Viewpointsystem GmbH, 1010 Wien, Austria}\\
\\
\normalsize{$^\ast$To whom correspondence should be addressed; E-mail: {vuletic@mit.edu}}
}

\begin{document} 


\baselineskip24pt


\maketitle 

\begin{sciabstract}
Quantum scrambling describes the spreading of local information into many degrees of freedom in quantum systems. This provides the conceptual connection among diverse phenomena ranging from thermalizing quantum dynamics to models of black holes.
Here we experimentally probe the exponential scrambling of a multi-particle system near a bistable point in phase space and utilize it for  entanglement-enhanced metrology. 
We use a time-reversal protocol to observe a simultaneous exponential growth of both the metrological gain and the out-of-time-order correlator, thereby experimentally  verifying the relation between quantum metrology and quantum information scrambling.   Our experiments demonstrate that fast-scrambling dynamics capable of exponentially fast entanglement generation are useful for practical metrology, resulting in 6.8(4) dB gain beyond the Standard Quantum Limit.
\end{sciabstract}

\setlength{\unitlength}{\textwidth}
\begin{figure*}[!htbp]
\centering
\includegraphics[width=.95\textwidth]{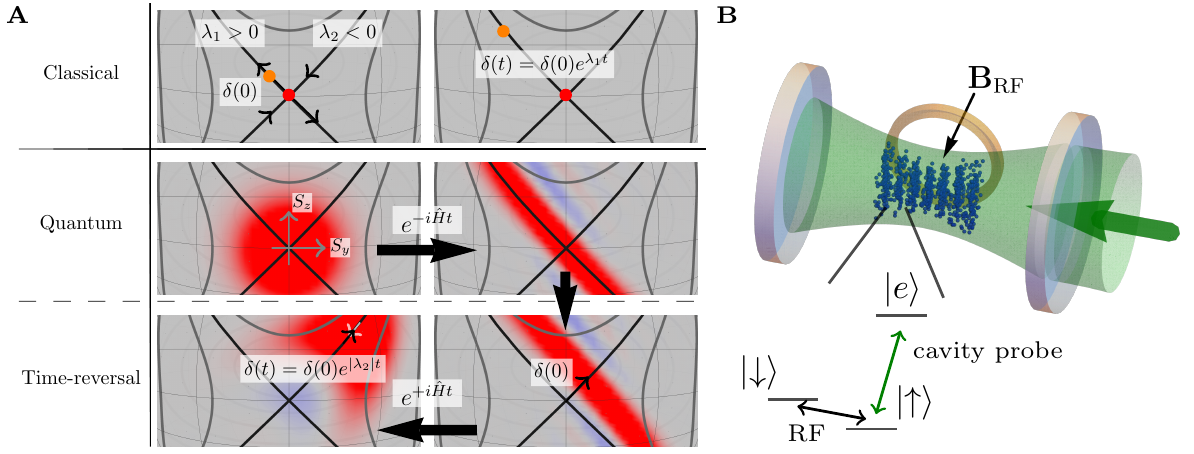}
\caption{\textbf{Time-reversal based exponential growth of sensitivity in a system with an unstable fixed point. a}, Classically, for a trajectory with a positive Lyapunov exponent $\lambda_1>0$ an initial signal (displacement) $\delta(0)$ increases exponentially over time. For quantum dynamics, however, an initial overlap between two states is preserved under unitary evolution. To amplify the signal similarly to the classical case, one needs to evolve the state under the nonlinear $\hat{H}$, resulting in decreased quantum fluctuations along a direction with negative Lyapunov coefficient $\lambda_2<0$. A displacement along this direction followed by application of the negative Hamiltonian $-\hat{H}$ (such that $\lambda_{1,2} \rightarrow -\lambda_{1,2}$) is then used to amplify the signal. {\bf b}, Experimental setup. The LMG Hamiltonian is generated by interaction of the collective atomic spin with light inside a cavity on the transition $\ket{\uparrow} \rightarrow \ket{e}$, while a radiofrequency magnetic field is applied to rotate the atomic spin. 
}
\label{fig:cartoon}
\end{figure*}

Even though every unitary dynamics of quantum systems is in principle reversible, it is extremely challenging in practice to reverse the arrow of time in generic interacting many-body systems. 
This is because any small perturbations or imperfections in time-reversed dynamics can lead to highly complicated, non-local changes in quantum wavefunctions, similar to the butterfly effect in chaos theory.
Dubbed information scrambling~\cite{Sekino2008, Swingle2016, Lewis2019}, this quantum mechanical effect gives rise to a variety of novel phenomena and applications ranging from models of traversable wormholes~\cite{Nezami2021quantum,Schuster2022} to quantum metrology~\cite{Appel2009}.
The speed of information scrambling is quantified by out-of-time-ordered correlators (OTOCs)~\cite{Larkin1969quasiclassical,KitaevTalk}. 
In certain systems, the OTOC grows exponentially fast over time $e^{\lambda_Q t}$, where $\lambda_Q>0$ defines the generalized quantum Lyapunov exponent~\cite{KitaevTalk}.
OTOCs have been measured~\cite{Li2017} and used as probes for various many-body phenomena, such as thermalization~\cite{Green2022}, quantum phase transitions~\cite{Wei2019}, many-body entanglement growth~\cite{Garttner2017}, and quantum scrambling~\cite{Landsman2019, Braumuller2022, Mi2021}.
However, the observation of exponential scrambling has remained elusive.

One approach to effective time-reversal involves 
changing the sign of the Hamiltonian $\hat{H}$$\rightarrow$$-\hat{H}$ during the evolution of highly engineered quantum systems. In the field of quantum metrology, it  enables a family of powerful quantum amplification protocols~\cite{Davis2016, frowis2016detecting, Hosten2016a, nolan2017optimal, macri2016loschmidt, Gilmore2021, Colombo2022,Mao2022Quantum} such as signal-amplification-through-time-reversed-interaction (SATIN)~\cite{Colombo2022}. Such protocols can be robust against many limitations that usually affect entanglement-enhanced atomic sensors, including imperfect measurements. In the case of exponentially scrambling dynamics 
(see Fig.~\ref{fig:cartoon}a), the SATIN signal as well is amplified exponentially over time. 

Here we experimentally implement a SATIN protocol for a Lipkin-Meshkov-Glick (LMG) Hamiltonian~\cite{lipkin1965validity, Duan2000,Hamley2012,Strobel2014,Muessel2015,Peise2015,Pilatowsky-Cameo2020,Muniz2020,Li2022,Mao2022Quantum} that exhibits exponential phase space evolution. The LMG Hamiltonian is generated in a cavity QED (cQED) system by adding a global rotation term $\hat{S}_x$ to the One-Axis-Twisting (OAT)~\cite{Kitagawa1993} Hamiltonian $\hat{S}_z^2$,
\begin{align}\label{eq:Hamiltonian}
    \hat{H}= \chi\hat{S}_z^2 - \Omega\hat{S}_x,
\end{align}
Here $\hat{\bf S}=(\hat{S}_x,\hat{S}_y,\hat{S}_z)$ represents the total spin of the system comprised of $N=2S$ spin-$\frac{1}{2}$ particles. 
While the time evolution is not chaotic due to the conservation of $\hat{S}^2$, the LMG Hamiltonian nevertheless features a quantum Lyapunov exponent for $0 < \Omega/(S \chi) <2$ due to an unstable (bifurcating) trajectory in the system phase space (see Fig. \ref{fig:cartoon}a)~\cite{Pappalardi2018,Rozenbaum2020,Pilatowsky-Cameo2020}.

\setlength{\unitlength}{\textwidth}
\begin{figure*}[!ht]
\centering
\includegraphics[width=.78\textwidth]{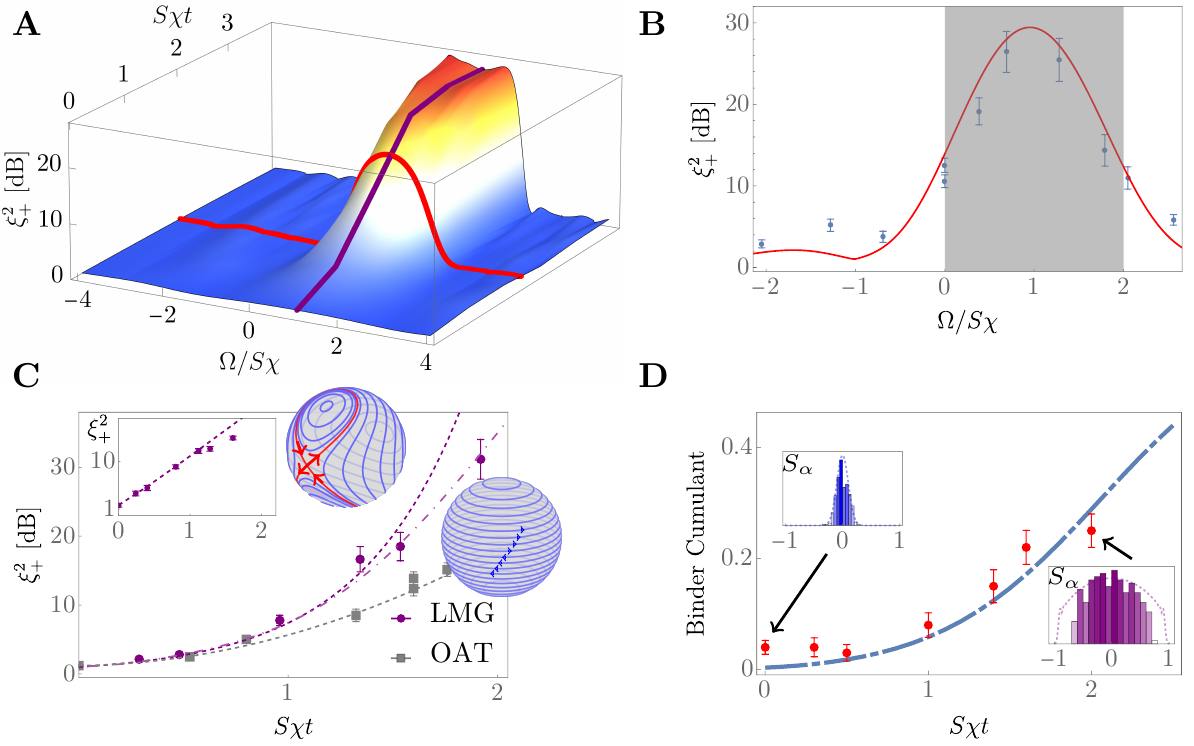}
\caption{\textbf{Collective-spin evolution in the CQED system. a,} 
Numerical calculation of the normalized variance $\xi_+^2$ of the antisqueezed direction as a function of $S\chi t$ and $\Omega/S\chi$ with linecuts representing the measurements in b, c. 
\textbf{b,}
The antisqueezing for a given $S\chi t=1.9$ as a function of the rotation strength $\Omega$. The shaded region indicates exponential growth, while in the other regions the time evolution is either quasi-periodic or growing polynomially. \textbf{c,}
Comparison of antisqueezing $\xi_+^2$ between the fastest exponential growth for a critical rotation strength $\Omega=S\chi$, and the polynomial growth of pure OAT ($\Omega=0$). The two Bloch spheres represent the lines of classical evolution in both situations. The dashed and dash-dotted red lines represent exponential growth based on the theoretical predicted Lyapunov exponent and the full numerical result, respectively. The gray dashed line is calculated for $\Omega=0$.
Inset: logarithmic plot for $\Omega=S\chi$ showing exponential growth of $\xi_+^2$.
\textbf{d,} The Binder cumulant, characterizing the shape of the distribution, for the antisqueezed direction for the critical LMG condition $\Omega=S\chi$ vs. time $t$. Insets: spin distribution with $S\chi t=0$ (blue) and $S\chi t=2$ (purple), with the latter being strongly non-Gaussian.
}
\label{fig:AntiSqueezing}
\end{figure*}

Our experiments operate with $N= 200$ \Yb atoms whose magnetic sublevels $\ket{\uparrow, \downarrow}$ in the electronic ground state represent a spin-$\frac{1}{2}$ system. One of the two spin states ($\ket{\uparrow}$) couples to an electronically excited state $\ket{e}$ via $\sigma^+$-polarized light that circulates inside the optical cavity (Fig. \ref{fig:cartoon}b). The coupling between a single atom and the cavity is characterized by the single-atom cooperativity $\eta=8.8(2)$~\cite{Tanji-Suzuki2011}.
We implement the LMG Hamiltonian in the rotating frame by adding an oscillating transverse magnetic field to the OAT Hamiltonian~\cite{Li2022} (see Fig. \ref{fig:cartoon}b and Supplementary Material (SM)).

The experiments start by initializing the system in a coherent spin state (CSS) pointing along the $x$-axis by means of optical pumping followed by a $\pi/2$ spin rotation. Analytical solutions using the Holstein-Primakoff approximation~\cite{Holstein1940} show that for $\Omega/(S\chi)<0$ or $\Omega/(S\chi)>2$ the system evolution is periodic with a frequency $\omega=\sqrt{\Omega^2+2S\chi\Omega}$~\cite{Law2001}. On the other hand, for $0< \Omega/(S\chi)<2$ the frequency $\omega$ becomes imaginary, corresponding to an unstable-fixed-point exponential evolution with a Lyapunov exponent $\lambda_Q=|\omega|$. For a fixed $S\chi$, choosing $\Omega=S\chi$ results in a maximum Lyapunov exponent $\lambda_Q=|S\chi|$.
At this specific parameter the LMG model for short time is an effective two-axis twisting (TAT) Hamiltonian $\hat{H}_\mathrm{TAT}=\chi\left(\hat{S}_z^2-\hat{S}_y^2\right)$~(see SM), which has been proposed for experimental implementation~\cite{Liu2011,Hamley2012} but not previously realized. 

We first measure the anti-squeezing (largest variance $\xi_+^2 \equiv \mathrm{max}_\alpha [\mathrm{var}(S_\alpha)]/(S/2)$) of the collective spin $\hat{\bf{S}}$ after an evolution under $\hat{H}$ as a function of the ratio $\Omega/(S\chi)$. The anti-squeezing $\xi_+$ constitutes an upper bound on the quantum Fisher information (QFI) with respect to spin rotations~\cite{Pezze2018}.
As shown in Fig. \ref{fig:AntiSqueezing}b, the experimental data for $\xi_+^2$ agree with the numerical simulation of the model (solid red line), and show a peak at $\Omega=S\chi$, as expected.

We then measure in Fig. \ref{fig:AntiSqueezing}c how $\xi_+^2$ grows with time for the two cases $\Omega =0$ (OAT Hamiltonian) and $\Omega = S\chi$ (critically tuned LMG Hamiltonian). 
The OAT data (gray) exhibit quadratic growth of $\xi_+^2$, as expected. 
The LMG data (red) show exponential growth $\xi_+^2 = e^{2\lambda_Q t}$ with $\lambda_Q=S \chi$ for times $t \lesssim (S\chi)^{-1}$. For larger times, the growths slows due to finite particle number and light-induced decoherence~\cite{Li2022} (see SM). The finite total spin further causes the states to turn non-Gaussian, which we characterize via the Binder cumulant~\cite{Binder1981finite}, as shown in Fig.~\ref{fig:AntiSqueezing}d. 

\setlength{\unitlength}{\columnwidth}
\begin{figure}[!htbp]
\centering
\includegraphics[width=\columnwidth]{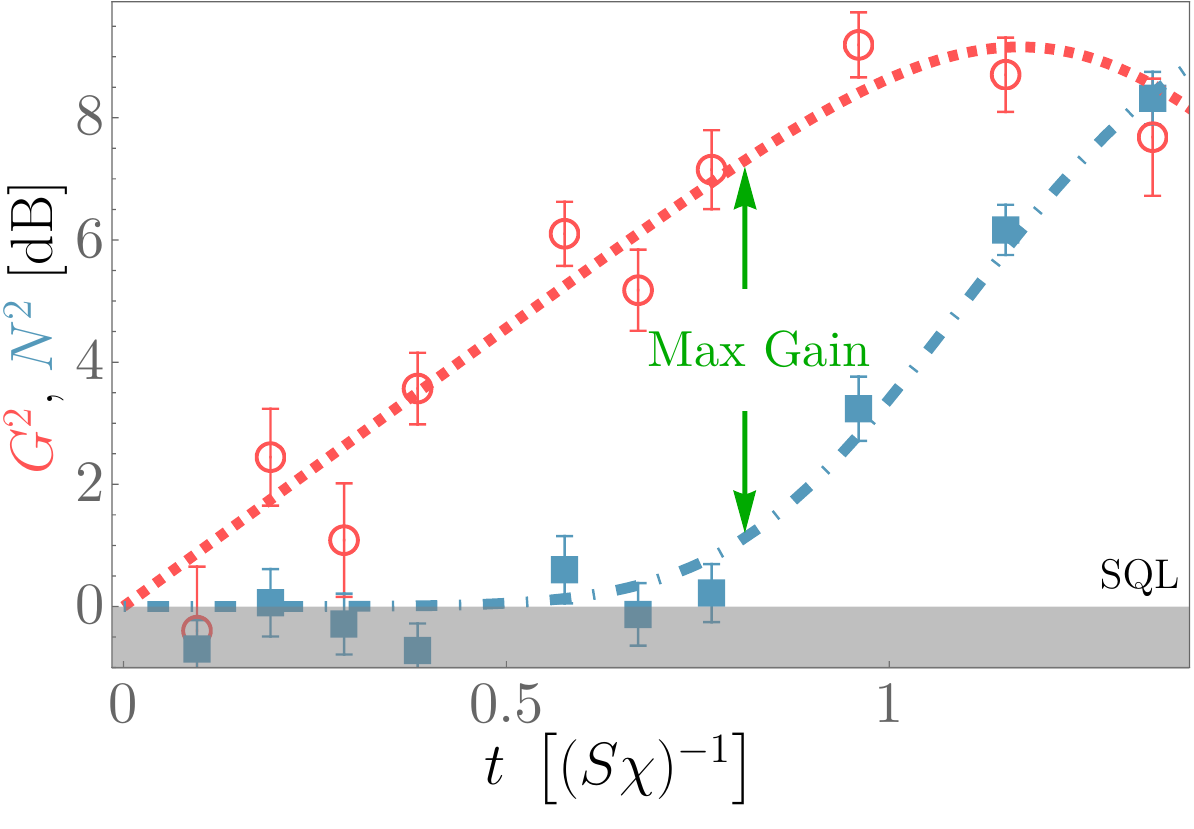}
\caption{\textbf{Metrological gain with exponential LMG time-reversal protocol.} 
The squared signal amplification $G^2$ (pink open circles) and system noise $N^2$ (blue solid squares) vs. time $t$. The orange dashed line represents the exponential growth of the anti-squeezing shown in Fig.~\ref{fig:AntiSqueezing}, representing an upper bound to the QFI. The blue dash-dotted line is the calculated noise due to residual light-atom entanglement. 
The maximum metrological gain is $6.8(4)$~dB. 
}
\label{fig:MetrologicalGain}
\end{figure}

The time evolution under the critically tuned ($\Omega=S\chi$) LMG Hamiltonian $\hat{H}$ quickly prepares an entangled collective quantum state.
To implement quantum metrology with the SATIN protocol, we then apply a small rotation
$\hat{U}_{\delta \phi} = e^{-i\hat{S}_\alpha\delta\phi}$, where $\hat{S}_\alpha \equiv \hat{S}_y \cos\alpha + \hat{S}_z \sin\alpha$ represents a collective spin operator in $yz$-plane.
This encodes a signal phase $\delta\phi$ along the $\alpha$ direction, with $\alpha=\pi/4$ chosen to maximize the metrological gain (see Fig. \ref{fig:cartoon}a and Supplementary Information).
To implement $-\hat{H}$, we switch to another set of laser frequencies incident on the cavity and flip the sign of the transverse field $\Omega$~(see Supplementary Information).
This generates an effective backward evolution in time that amplifies the applied signal $\delta\phi$. The shifted state then undergoes a bifurcated trajectory for $\delta \phi \lessgtr 0$ (see Fig. \ref{fig:AntiSqueezing}c), and results in an exponentially amplified deviation $G \delta \phi$ from the original position. 
As shown in Fig.~\ref{fig:MetrologicalGain}, the squared signal amplification $G^2$ (orange) increases exponentially with the same exponent $2\lambda_Q$ as the anti-squeezing $\xi_+^2$ up to times $t \approx (S\chi)^{-1}$.
The the measured quantum noise $N^2$, i.e. the variance of spin projection noise along the amplification direction $\hat{S}_\alpha$ normalized to the standard quantum limit (SQL) (blue) remains unity until $t \approx 0.8(S\chi)^{-1}$. The increase of the noise $N^2$ results from the residual light-atom entanglement~\cite{Li2022}, and can be improved in the future by optimizing the light detuning (see SM).
The improvement of the metrological gain over the SQL is 6.8(4)~dB. The deviation of $G^2$ from an exponential for $t \gtrsim (S\chi)^{-1}$ is due to the non-uniform coupling between atoms and the cavity light~\cite{Hu2015} as well as the residual light-atom entanglement, both of which can be improved in the future~\cite{Wu2021,Li2022}.

\begin{figure}[!tbp]
    \centering
    \includegraphics[width=\columnwidth]{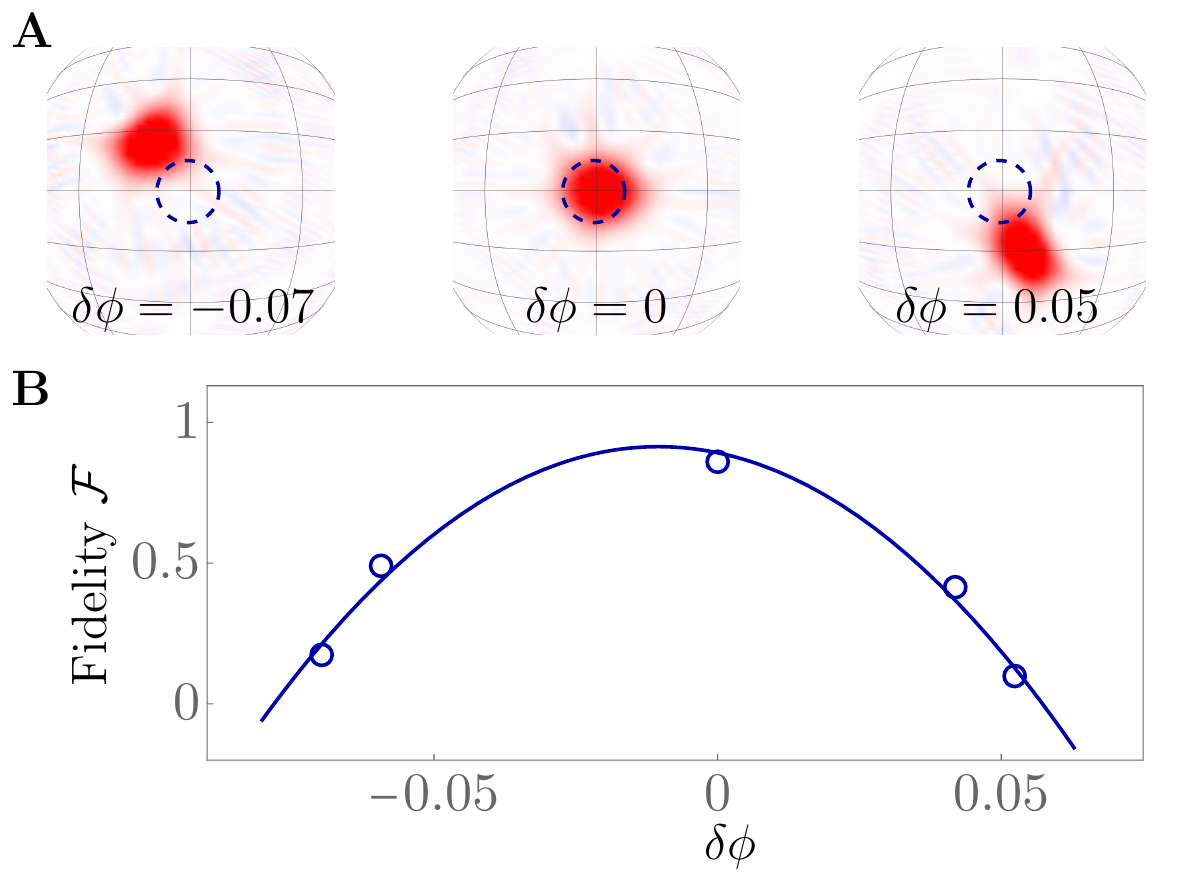}
    \caption{\textbf{FOTOC and OTOC extracted from quantum state tomography.} {\textbf a,} Experimental Wigner functions obtained from quantum state tomography after the LMG SATIN protocol with different signal displacements $\delta\phi$ (for $\Omega=S\chi$ and $t=0.57(S\chi)^{-1}$).
    The dashed circle indicates the orginal CSS state. {\textbf b, }
    The solid blue line is a quadratic fit used to extract the OTOC $\mathcal{I}$ (see text and eq. \ref{eq:OTOC}).
    }
\label{fig:tomo}
\end{figure}

To investigate the QIS aspect of the time-reversal protocol, we measure the FOTOC with quantum state tomography using randomized measurements~\cite{Schmied2011,randomizedReview} (see SM). 
The FOTOC $\mathcal{F}(t)$ can be expressed as the trace between the density matrix $\rho(0)$ of the original state and that of the state displaced by $\delta \phi$ evolved backward in time, $\rho_t'(0)=\hat{U}_t \hat{\rho}(0) \hat{U}_t^{\dagger}$, where $\hat{U}_t = e^{i\hat{H}t} e^{-i\hat{S}_\alpha\delta\phi} e^{-i\hat{H}t}$,
\begin{align}
\begin{aligned}
\mathcal{F}(t) \equiv\left\langle\hat{U}_t \hat{\rho}(0) \hat{U}_t^{\dagger} \hat{\rho}(0)\right\rangle =\operatorname{Tr}\left(\rho_t^{\prime}(0) \rho(0)\right).
\end{aligned}
\end{align}
At fixed forward evolution time $t$, the FOTOC $\mathcal{F}$ depends on the small displacement $\delta \phi$ and relates to the OTOC $\mathcal{I}(t)$ by its second derivative~\cite{Wei2019}
\begin{align}
    \mathcal{I}(t)\equiv -\frac{1}{2}\left.\frac{\partial^2\mathcal{F}(t)}{(\partial \delta\phi)^2}\right|_{\delta\phi = 0}=\left\langle\hat{S}_\alpha(t) \hat{\rho}(0) \hat{S}_\alpha(t) \hat{\rho}(0)\right\rangle, 
    \label{eq:OTOC}
\end{align}
with the Hermitian operator $\hat{S}_\alpha(t)\equiv e^{i\hat{H}t}\hat{S}_\alpha e^{-i\hat{H}t}$.

Choosing four different evolution times (such that $S\chi t_1 \in \{0.38, 0.57, 0.77, 0.96\}$), we displace the entangled state for each $t_1$ by several different small angles $\delta\phi$. We then perform the tomographic reconstruction after a reversed time evolution with $-\hat{H}$ to obtain $\mathcal{F}(t_1)$, as shown in Fig.~\ref{fig:tomo}(a). The OTOC $\mathcal{I}(t_1)$ is then extracted from the data by fitting a quadratic function in the displacement $\delta \phi$ to the FOTOC (Fig. \ref{fig:tomo}b). We notice that the fitted quadratic curve is slightly shifted from $\delta \phi=0$, and has slightly reduced peak fidelity. 
The shift is likely due to a small difference between the assumed and the real Larmor frequencies between the spin states, while the reduction from unit peak fidelity is due to the imperfect time reversal associated with residual light-atom entanglement. The small imperfections do not reduce the metrological gain significantly~(see SM). 

\begin{figure}[!tbp]
    \centering
    \includegraphics[width=\columnwidth]{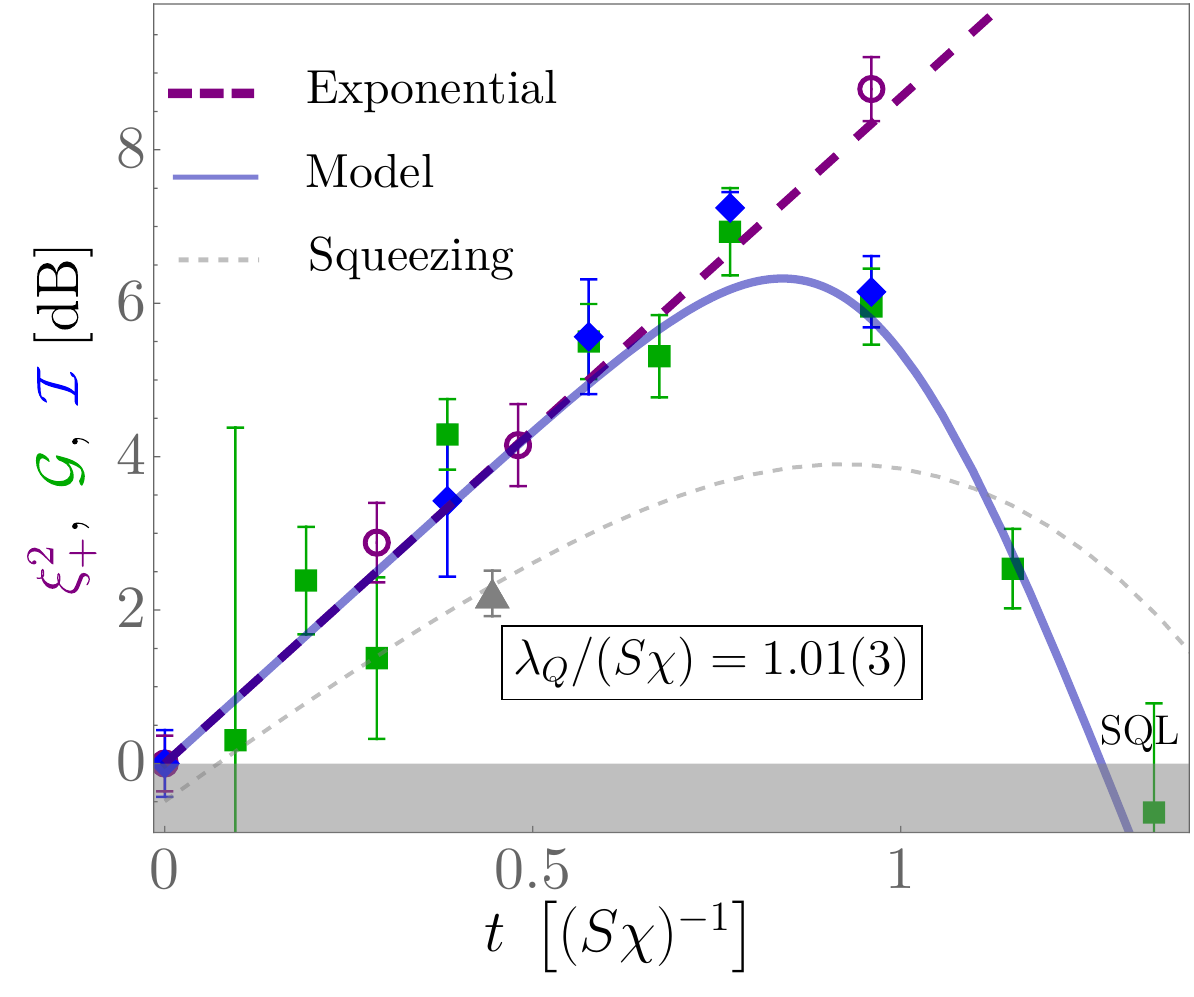}
    \caption{{\bf Comparison between quantum information and quantum metrology parameters for the LMG model.} The red open circles, green solid squares and blue solid diamonds represent the antisqueezing, metrological gain, and OTOC, respectively. All quantities increase initially exponentially with time with a fitted Lyapunov exponent $\lambda_Q=1.01(3)S\chi$ that agrees well with the theoretical prediction $\lambda_Q=S\chi$. The OTOC errorbars are obtained by using the bootstrapping method~\cite{hardle1991bootstrap} (see SM). For longer times $t\gtrsim (S\chi)^{-1}$, the metrological gain and OTOC decrease due to decoherence caused by light-atom entanglement, as is well captured by the theoretical model (blue solid line). The gray dashed line (gray data point) represent the calculated (measured) metrological gain using the squeezing generated by the LMG Hamiltonian. 
    }
    \label{fig:final_comparison}
\end{figure}

Figure~\ref{fig:final_comparison} summarizes our findings regarding the close relation between quantum scrambling and time-reversal quantum metrology: The antisqueezing $\xi_+^2$, metrological gain $\mathcal{G}$, and OTOC $\mathcal{I}$ all agree with each other and scale exponentially with application time $t$ of the LMG Hamiltonian for $t \lesssim 0.8(S \chi)^{-1}$. The exponential fit yields a Lyapunov exponent $\lambda_Q/(S\chi) = 1.01 \pm 0.03$, in excellent agreement with the theoretical expectation $\lambda_Q/(S\chi) = 1$.

Our experiments demonstrate a CQED realization of the critically tuned ($\Omega = S \chi$) LMG model with an exponential evolution in phase space. Besides, we also point out and experimentally verify that time reversal protocols represent a powerful experimental tool giving access not only to metrological gain beyond the SQL~\cite{Gilmore2021,Colombo2022,barberena2022fast}, but also enabling the measurement of quantum information scrambling in large many-body systems. We observe exponential growth of both the OTOC and the metrological gain for the LMG model, thereby experimentally verifying the intrinsic relation between these two concepts from different subfields of quantum science. 
The demonstrated methods to reverse time may enable the experimental investigation of complex many-body quantum systems where the information spreads exponentially fast within many degrees of freedom, relevant for the simulation of black holes and quantum gravity models in controlled table-top experiments~\cite{Sekino2008}. 
In addition, we envisage that time-reversal protocols will readily render many fast-scrambling Hamiltonians useful for robust and fast quantum-enhanced metrology.

\section*{Acknowledgements}

We thank James Thompson, Monika Schleier-Smith, Boris Braverman and Albert Adiyatullin for inspiring discussions. 



Competing interests: The authors declare no competing interests.

\bibliographystyle{Science}
\bibliography{SWRLMG}
\newpage
\setcounter{section}{0}
\section*{Supplementary Material}
\def\thesubsection{\thesection-\Alph{subsection}}



\subsection*{Loading}
We load $^{171}$Yb atoms from a Yb dispenser into a two-color mirror magneto-optical trap (MOT) on the singlet ${^1S_0}{\rightarrow} {^1P_1}$ and triplet ${^1S_0}{\rightarrow} {^3P_1}$ transitions.
In this stage, the atoms are cooled down to $\approx 1$~mK, close to the Doppler limit imposed of the singlet transition.
Subsequently, the are loaded into a second-stage green MOT on the narrower-linewidth triplet transition that allows for a compressed ($\approx 100~\mathrm{\mu m}$) and colder atomic cloud ($\approx 20 \mathrm{\mu K}$).
By moving the zero-magnetic-field point, the atomic cloud is transported into the cavity TEM$_{00}$ mode at a distance of $\Delta z \approx 50 \mathrm{~\mu m}$ from the cavity waist~\cite{Kawasaki2020}. 

The trap is formed, through the cavity, by standing-wave light at the `magic-wavelength' $\lambda_t\approx759$~nm.
At the atom location, the trap depth is $U_0 = k_B~120\times\mathrm{\mu K}$.
Since the trapping light is always on, simply turning off the green MOT transfers the atoms into the optical lattice.
However, before turning of the green MOT light, we reduce the light intensity and simultaneously increase the red detuning of the frequency; this last step allows us to compensate for the differential Stark shift induced by the trapping light which on the triplet state ${}^3P1$ is $25\%$ larger than on the ground state.   

In this way, an ensemble of atoms is prepared at a distance of $\approx 50$~$\mu$m from the cavity mode waist.

\subsection*{State measurement}
%
The single-shot measurement outcome of $\hat{S}_z$ is obtained from the difference ${S}_z{=}({N}_\uparrow{-}{N}_\downarrow){/}2$ of the populations $N_\uparrow$ and $N_\downarrow$ of the states $\ket{\uparrow}$ and $\ket{\downarrow}$, respectively. 
We first measure the $N_\uparrow$ through the vacuum Rabi splitting of the cavity mode $2g{\approx}\sqrt{N_\uparrow\eta\kappa\Gamma}$ when the empty cavity is resonant with the transition $\ket{\uparrow}\rightarrow\ket{e} \equiv \ket{^3P_1, m_F=+3/2}$~\cite{Braverman2019}. 
Here $\kappa{=}2\pi{\times}530(10)~$kHz is the cavity linewidth and $\Gamma{=}2\pi{\times}184(1)~$kHz the linewidth of the atomic transition. The Rabi splitting is measured by scanning the laser frequency and detecting the cavity transmission as a function of the frequency. To measure $N_\downarrow$ we swap the population between the $\ket{\uparrow}$ and $\ket{\downarrow}$ states through the application of an rf $\pi$-pulse and repeat then the measurement procedure described above. The sum $N_\uparrow + N_\downarrow$ gives the total atom number $N$ in one realization of the experiment.  

The resolution of a single measurement, normalized to the SQL, is given by $\sigma_d^2= \frac{1}{S_0}\mathrm{var}(S_{z1}-S_{z2})$ where $S_{z1}$ and $S_{z2}$ are two state measurements performed after a single coherent spin state (CSS) preparation. We obtain $\sigma_d^2{=}0.15\pm0.02$, and it remains constant within the whole range of atom numbers used in this experiment. 
For experiments requiring a measurement of any other projection of the spin vector, we apply a set of dedicated rf-pulses that map the desired projection onto $\hat{S}_z$.

\subsection*{Characterizing the single-atom cooperativity}

We characterize the single-atom cooperativity $\eta$ by measuring the spin projection noise via the cavity as a function of the collective cooperativity $N\eta$, where $N$ is the total number of atoms coupled to the cavity.
For a CSS prepared at the equator of the generalized Bloch sphere, the measured variance of the difference is 
\begin{equation}
\mathrm{var}(\eta S_z)= \frac{1}{4} N\eta^2.
\end{equation}
The latter contribution is obtained through the variance of the difference between two measurements after a single CSS preparation.

\begin{figure}[!tbp]
\centering
\includegraphics[width=.8\columnwidth]{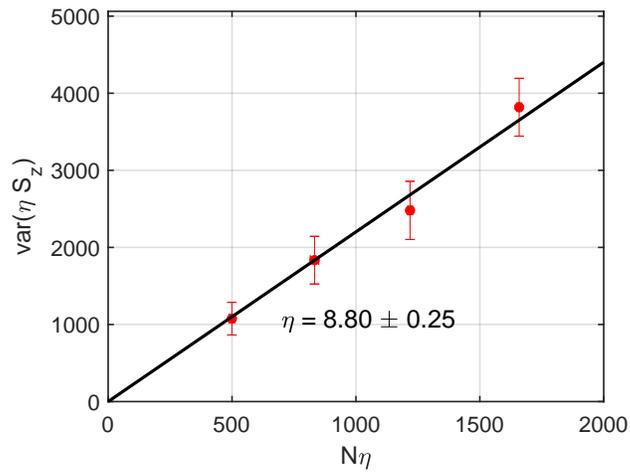}
\caption{Measured variance of $\eta S_z$ as a function of the average collective cooperative $N\eta$. When we are spin-projection limited, we have a linear relation with slope $\eta/4$.}
\label{fig:singleatomCooperativity}
\end{figure}


Plotting the variance of $\eta S_z$ as a function of the collective cooperativity $N \eta$ results, in the absence of classical sources of noise, in a linear graph with slope $\eta/4$ (see fig.~\ref{fig:singleatomCooperativity}).
We subtract the measurement resolution contribution to the variance and obtain single-atom cooperativity $\eta = 8.8\pm0.3$ by fitting the data to a linear model. 

When a quadratic fitting term is included to account for possible technical noise, we recover the same cooperativity $\eta=8.2 \pm 1.1$ within error bars. Moreover, the coefficient of the second-order term agrees with 0. 

\subsection*{Magnetic Rabi frequency}
We can rotate the collective spin states via $\hat{H}_\mathrm{trans}=\Omega {\bf n}\cdot\hat{\bf S}$, where $\hat{\bf S}$ is a vector operator of the collective spin-state with total length $S$, while its $x, y, z$-components are defined as $\hat{S}_{x,y,z} = \frac{1}{2} \sum_{j} \hat{\sigma}_{x,y,z}^j$. We can control the magnetic field direction ${\bf n}$ by controlling the RF field's phase and frequency. In this manuscript, we don't implement any $\hat{S}_z$ rotation; thus, controlling the phase is enough to tune ${\bf n}$ in the $x-y$ plane. 

We characterize our maximal magnetic field strength $\Omega_\mathrm{max}$ by a Rabi sequence.
We can generate a maximum Rabi frequency of $\Omega_\mathrm{max}=2\pi\times220$Hz.

\subsection*{Squeezing strength}
The squeezing strength is characterized by the anti-squeezing under a pure OAT Hamiltonian, which scales as $\mathrm{var}(\hat{S}_y)=1+\dot{F}t+(N\chi t)^2$. By fitting we obtain both the measurement rate $\dot{F}$ and the squeezing strength $N\chi$, whose ratio agrees with the theoretical calculation using parameters in Table~\ref{tab:my_label}.

\subsection*{Two-color Interaction}

The atoms are trapped by an optical lattice formed by the cavity mode at 759~nm, and the cavity light which generates the OAT Hamiltonian is at 556~nm. Thus, atoms trapped at different lattice site may experience different atom-light interaction strength (cooperativity). The collective behavior can be described by an effective atom number and effective cooperativity~\cite{Hu2015}; however, to obtain the ideal Hamiltonian, we need to cancel the first-order phase shift term $\hat{H}_0\propto\hat{S}_z$~\cite{Li2022}. We cannot use a detuning in the RF field to cancel that because each atom experiences a different local light field, and therefore a combination of different detunings must be applied. 

We simultaneously send two frequencies of light into the cavity, both generating the same squeezing sign but with opposite light-shift terms; the linear $S_z$ term of the Hamiltonian is then canceled by carefully tuning the amplitude of the two sidebands. The idea is similar to our previous setups~\cite{Braverman2019,Colombo2022}; the difference is that now we simultaneously send both sidebands to the cavity.
In this case, the full Hamiltonian~\cite[Eq.~(14)]{Li2022}'s electric field is the sum of two components with different frequencies, which individually generate the OAT and opposite first-order phase shifts. 

The Hamiltonian also contains an interference term which oscillates at the difference frequency of several MHz. This frequency is much higher than all other relevant frequency scales and has no significant effect. The frequencies chosen for $N\eta=2000$ and $\eta\sim 9$ are given in Table.~\ref{tab:my_label}.

\begin{table}[!ht]
    \centering
    \begin{tabular}{c|c|c}
         & Frequency 1 [MHz] & Frequency 2 [MHz] \\\hline
        Squeezing ($\chi>0$) & $8.93$ & $-2.19$ \\\hline
        Un-Squeezing ($\chi<0$) & $-8.9$ & $2.25$ \\\hline
    \end{tabular}
    \caption{Frequencies for two-color interaction.}
    \label{tab:my_label}
\end{table}


\subsection*{Hamiltonian}
The Hamiltonian for the collective spin consists of two parts, the OAT Hamiltonian and the transverse Rabi field:
\begin{align}
    \hat{H}=\hat{H}_\mathrm{OAT}+\hat{H}_\mathrm{trans}=\chi\hat{S}_z^2+\Omega{\bf n}\cdot\hat{\bf S}.
\end{align}
In our system, all parameters are controlled as described above. 

We now show that for short times, the LMG Hamiltonian with $\chi=\Omega/S$ and ${\bf n}=-\hat{\bf e}_x$ 
(i.e. the critical LMG condition in the main text) generates the same dynamics as the Two-Axis Twisting (TAT) Hamiltonian. By assuming that the initial coherent state is centered at $\ket{S_x=S}$, we can apply the Holstein-Primakoff approximation \cite{Holstein1940} to replace the collective spin operators by bosonic mode operators $\hat{a}$ as
\[\hat{S}_x=\frac{N}{2}-\hat{a}^\dagger \hat{a}, \hat{S}_y=\frac{\sqrt{N}}{2i}(\hat{a}^\dagger - \hat{a}),  \hat{S}_z=\frac{\sqrt{N}}{2}(\hat{a}^\dagger + \hat{a}), \]
and the LMG Hamiltonian reduces to
\[\hat{H}\sim\chi \frac{N}{4}\left(\hat{a}^{\dagger\,2}+\hat{a}^2\right) + \mathrm{const.}\]
The latter has exactly the form of the TAT Hamiltonian $\hat{H}_\mathrm{TAT}=(\hat{S}_y^2-\hat{S}_z^2)\sim-\frac{N}{2}(\hat{a}^{\dagger\,2}+\hat{a}^2)$.

\subsection*{Quantum state tomography and Fidelity OTOC}

We implement a quantum state tomography on the symmetric  Hilbert space's subspace of collective spin states~\cite{Schmied2011}. For each quantum state, we measure the spin projection along 41 different directions on the Bloch sphere with in total more than 1200 measurements, and reconstruct the state. Each measurement result is translated into a projection operator and fed into the algorithm to reconstruct a density matrix $\rho$. With the reconstructed density matrix, we can obtain the fidelities between them to get the fidelity OTOCs. 

\end{document}